\documentclass[12pt]{iopart}
\usepackage{graphicx}
\begin{document}
\title{Two-size Probability-Changing Cluster Algorithm} 
\author{Tasrief Surungan$^{1}$ and Yutaka Okabe$^{2}$}
\address{
$^1$Department of Physics, Hasanuddin University, Makassar, 
South Sulawesi 90245, Indonesia \\
$^2$Department of Physics, Tokyo Metropolitan University, Hachioji, 
Tokyo 192-0397, Japan 
}
\ead{tasrief@unhas.ac.id; okabe@phys.se.tmu.ac.jp}
\vspace{10pt}
\def\l{\langle}
\def\r{\rangle}

\date{\today}

\begin{abstract}
We propose a self-adapted Monte Carlo approach to automatically 
determine the critical temperature  by simulating 
two systems with different sizes at the same temperature. 
The temperature is increased or decreased by checking 
the short-time average of the correlation ratios of the two system sizes. 
The critical temperature is achieved using 
the negative feedback mechanism, and the thermal average 
near the critical temperature can be calculated precisely. 
The proposed approach is a general method to treat 
second-order phase transition, first-order phase transition, 
and Berezinskii-Kosterlitz-Thouless transition on the equal footing. 
\end{abstract}

\maketitle

\section{Introduction}

Finite-size scaling (FSS) \cite{fisher70} is a basic concept in the 
study of phase transitions and critical phenomena. The Binder ratio 
\cite{Binder81}, essentially the moment ratio, 
is widely used in the analysis of the numerical data. 
The moment ratios of the magnetization $m$, 
\begin{equation}
   U(T) = \l m(T)^4 \r/\l m(T)^2 \r^2,
\end{equation}
for different sizes scale as
\begin{equation}
   U(T) = f(t L^{1/\nu}),
\label{FSS}
\end{equation}
where $t=(T-T_c)/T_c$ and $\nu$ is the correlation-length exponent.
The linear system size is denoted by $L$. 
We can determine the critical temperature $T_c$ of the second-order 
phase transition by employing a condition where $U(T)$ does not 
depend on $L$.  By measuring $U(T)$ for different sizes, we can determine 
$T_c$ from the crossing point of temperature-dependent curves 
of different sizes.  We note that there are corrections 
to FSS.  There are other quantities that satisfy the scaling 
form such as Eq.~(\ref{FSS}); the second-moment correlation length 
divided by $L$ \cite{katzgraber} and the ratio of the correlation functions 
with different distances \cite{tomita2002b} are examples of such quantities. 

Moreover, if we consider the ratio of $U(T)$ with different sizes, 
e.g., $U(T,L)/U(T,L/2)$, the critical value of this ratio becomes one,
even if the critical value of $U(T)$ itself is not a universal one, 
and does depend on the model.  The ratios of $U(T)$ 
with different sizes were studied in the analysis of 
the Potts model \cite{Caraccido,Salas}. 
These ratios were also used in the recent analysis 
of the clock model \cite{Surungan}.

The FSS analysis is often associated with the Monte Carlo simulation. 
To overcome the slow dynamics in the single-spin flip 
algorithm, a multi-cluster flip algorithm was proposed 
by Swendsen and Wang \cite{sw87}.  
Wolff \cite{wolff89} proposed another type of cluster algorithm, 
that is, a single-cluster flip algorithm. 
Tomita and Okabe \cite{PCC} developed a cluster algorithm,
called the probability-changing cluster (PCC) algorithm,
for automatically determining the critical point.  
It is an extension of the cluster algorithm, 
but it changes the probability of cluster update 
(essentially, the temperature) during the Monte Carlo process.  

This paper presents a self-adapted method using two system sizes 
for automatically determining the critical temperature, 
which is referred to as the two-size PCC algorithm.  
We simultaneously perform the Monte Carlo simulations 
for the two system sizes. We measure 
some quantity $U$, which follows the scaling form 
shown in Eq.~(\ref{FSS}), and calculate the ratio of 
$U(L)/U(L/2)$ for short time.  Then, we increase or 
decrease the temperature by checking the value of $U(L)/U(L/2)$. 

We start with the two-size PCC algorithm for the second-order 
transition.  As an example, we treat the two-dimensional (2D) Ising model, 
and demonstrate how the critical temperature can be determined 
in a self-adapted way.  We calculate the thermal average of 
the physical quantities near the critical temperature. 
We also study the first-order transition. 
As a typical example, we deal with the 2D 6-state Potts model. 
We investigate the first-order transition temperature 
and the latent heat.  We demonstrate that the same procedure 
is also effective for studying the Berezinskii-Kosterlitz-Thouless 
(BKT) transition, where a fixed line instead of a fixed point exists. 
We select the 2D 5-state clock model that has two BKT 
transitions with higher and lower transition temperatures. 

The remaining part of the paper is organized as follows: 
We describe the two-size PCC algorithm for the second-order 
transition in Section II. 
Two-size PCC studies of the first-order transition 
and BKT transition are discussed in Sections III and IV, 
respectively. 
Section V is devoted to summarizing the study and discussing results.  

\section{Second-order transition}
\begin{figure}
\begin{center}
\includegraphics[width=0.54\linewidth]{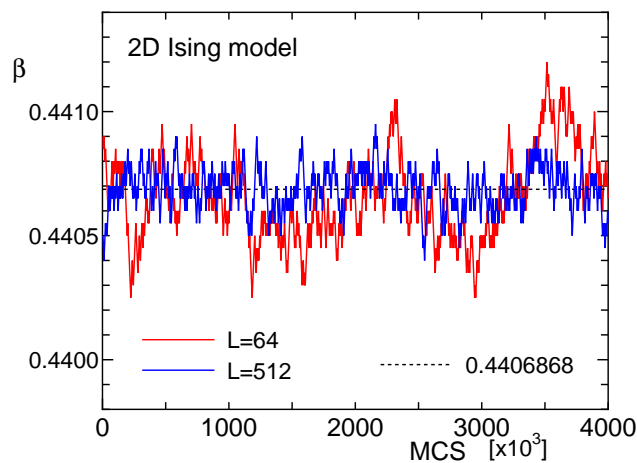}
\caption{
The time evolution of $\beta$ with the two-size PCC algorithm 
for the 2D Ising model. 
The system sizes are $L = 64$ and $L = 512$; 
that is, the set of system sizes are (64,32) and (512,256). 
The exact value of $\beta_c$ (=$\ln(1+\sqrt{2})/2=0.4406868$) 
for an infinite system is denoted by a dotted line.
}
\label{fig:time_Ising}
\end{center}
\end{figure}


Let us start with the second-order phase transition. 
As an example, we consider a 2D Ising model on the square lattice, 
whose Hamiltonian is given by
\begin{equation}
 H = -J \sum_{\l ij \r} \sigma_i \sigma_j, \quad \sigma_i = \pm 1. 
\label{Ising}
\end{equation}
The summation is taken over the nearest-neighbor pairs, 
and periodic boundary conditions are imposed in numerical 
simulations. 
Using the Wolff single-cluster flip algorithm 
for spin update, we simulate the two system sizes simultaneously. 
In determining the  critical point (line), we use the 
ratio of the correlation functions with different distances, 
i.e., the correlation ratio \cite{tomita2002b}, 
\begin{equation}
  R(T) = \l g(r) \r/\l g(r') \r. 
\label{corr_ratio}
\end{equation}
Here, $g(r)$ is a correlation function with the distance $r$. 
For the values of $r$ and $r'$, we choose $r=L/2$ and $r'=L/4$ 
for numerical calculations.

The actual procedure for the two-size PCC algorithm is as follows. 
We use two systems of different sizes, 
say $L$ and $L/2$.  After simulating some steps
at the same temperature, we measure the correlation ratios 
of both the systems, $R(L)$ and $R(L/2)$. 
We increase or decrease the inverse temperature $\beta$ 
(= $1/T$ in units of the coupling $J$) according
to the following rule: 
\begin{eqnarray}
\label{feedback}
\beta = \left\{
  \begin{array}{ll}
    \beta + \Delta \beta & \quad \mbox{if $R(L)/R(L/2) < 1$}, \\
    \beta - \Delta \beta & \quad \mbox{otherwise},
  \end{array}
  \right.
\end{eqnarray}
where $\Delta \beta > 0$.
There are two parameters to choose, the number of Monte Carlo steps (MCS) 
for taking a short-time average, $N_{\rm av}$, and the difference 
of $\beta$, $\Delta \beta$. 
Note that the cluster flip algorithm is effective 
because the rapid equilibration is required after a change in temperature. 

The plot of the time evolution of $\beta$ is shown 
in Fig.~\ref{fig:time_Ising}. 
For this plot, we chose $N_{\rm av}$ = 4000 and $\Delta \beta$ = 0.00005; 
that is, after every 4000 MCS, $\beta$ is changed by $\pm 0.00005$. 
We will discuss the choice of $N_{\rm av}$ and $\Delta \beta$ later. 
In the figure, the time steps are given in units of 1000 MCS, 
and we show the data up to $4 \times 10^6$ MCS.  
The system sizes are $L = 64$ and $L = 512$; 
that is, the set of system sizes are (64,32) and (512,256). 
We see that the temperature oscillates 
around the average value. 
The width of fluctuation decreases as the system size increases 
because of the effect of self-averaging. 
For convenience, we denote the exact value of $\beta_c$ 
(=$\ln(1+\sqrt{2})/2=0.4406868$) for an infinite system 
by a dotted line.

\begin{figure}
\begin{center}
\includegraphics[width=0.54\linewidth]{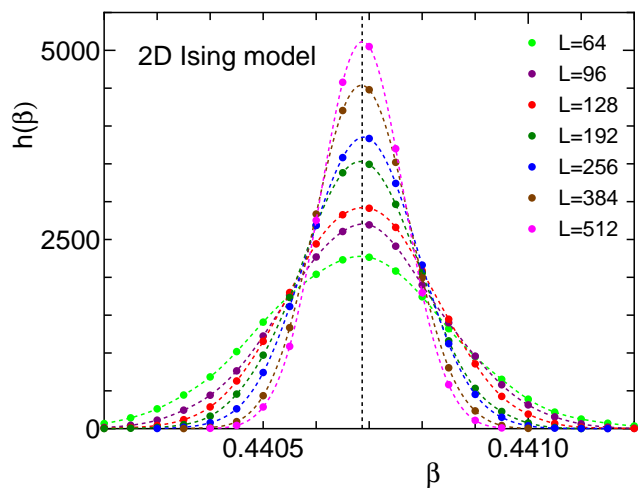}
\caption{
The size dependence of the histogram of $\beta$, $h(\beta)$, 
for the 2D Ising model. 
The system sizes are $L$ = 64, 96, 128, 192, 256, 384, and 512; 
the condition for averaging is
$N_{\rm av}$=4000, $\Delta \beta$=0.00005.
The exact value of $\beta_c$ (=$\ln(1+\sqrt{2})/2=0.4406868$) 
for an infinite system is denoted by a dotted line.
}
\label{fig:h(b)i}
\end{center}
\end{figure}

Next, we examine the histogram of $\beta$, $h(\beta)$, 
for the two-size PCC algorithm. 
We plot $h(\beta)$ of the 2D Ising model for $L$ = 64, 96, 128, 192, 
256, 384, and 512 in Fig.~\ref{fig:h(b)i}. 
Measurement is performed for $4 \times 10^6$ MCS 
after equilibration of 10,000 MCS. We made 32 runs 
for each system size in order to estimate statistical errors. 
For smaller sizes ($L$ = 64, 96, and 128), we made 64 runs. 
The parameters $N_{\rm av}$ and $\Delta \beta$ were 
chosen as 4000 and 0.00005, respectively. 
In the plot, the histogram $h(\beta)$ is normalized by
$$
   \int h(\beta) \ d\beta = 1.
$$
The obtained histogram is similar to normal distribution, 
and we see that the histogram becomes sharper 
with an increase in system size. Furthermore, the peak position 
approaches the exact value of $\beta_c$ for the infinite system. 

\begin{figure}
\begin{center}
\includegraphics[width=8.6cm]{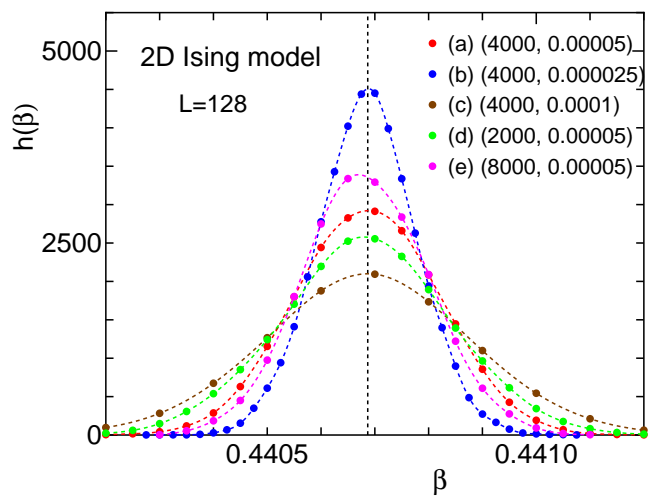}
\caption{
The comparison of the histogram of $\beta$, $h(\beta)$, 
for the 2D Ising model. The system sizes is set to be $L$ = 128, 
and the conditions for averaging are 
(a) $N_{\rm av}$=4000, $\Delta \beta$=0.00005, 
(b) $N_{\rm av}$=4000, $\Delta \beta$=0.000025, 
(c) $N_{\rm av}$=4000, $\Delta \beta$=0.0001,  
(d) $N_{\rm av}$=2000, $\Delta \beta$=0.00005, 
(e) $N_{\rm av}$=8000, $\Delta \beta$=0.00005. 
}
\label{fig:h(b)128i}
\end{center}
\end{figure}

Here, we examine the choice of $N_{\rm av}$ and $\Delta \beta$. 
Figure~\ref{fig:h(b)128i} shows a comparison of 
$h(\beta)$ of the 2D Ising model 
with $L=128$ for five conditions; 
(a) $N_{\rm av}$=4000, $\Delta \beta$=0.00005, 
(b) $N_{\rm av}$=4000, $\Delta \beta$=0.000025, 
(c) $N_{\rm av}$=4000, $\Delta \beta$=0.0001,  
(d) $N_{\rm av}$=2000, $\Delta \beta$=0.00005, 
(e) $N_{\rm av}$=8000, $\Delta \beta$=0.00005. 
The histogram becomes sharper with a decrease in $\Delta \beta$ 
and an increase in $N_{\rm av}$. 
The systematic size dependence is obtained 
when the conditions of $N_{\rm av}$ and $\Delta \beta$ are fixed. 
In the following, we will show the data for condition 
(a) $N_{\rm av}$=4000, $\Delta \beta$=0.00005. 

\begin{figure}
\begin{center}
\includegraphics[width=8.6cm]{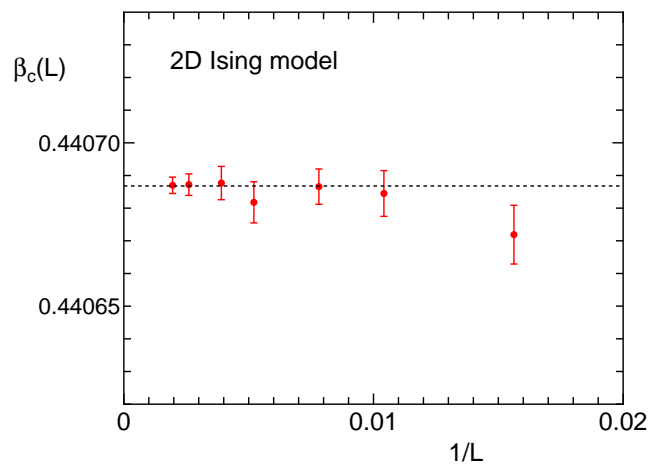}
\caption{
The plot of $\beta_c(L)$ 
of the 2D Ising model. 
The linear system sizes $L$ are 64, 96, 128, 192, 256, 384, and 512. 
The exact 
value is denoted by a dotted line.
}
\label{fig:betac_Ising}
\end{center}
\end{figure}

The transition (inverse) temperature for each size, 
$\beta_c(L)$, was estimated using the averaged value of $\beta$. 
The plot of the size dependence of $\beta_c(L)$ as a function of $1/L$ 
is shown in Fig.~\ref{fig:betac_Ising}, 
where the statistical errors were estimated by $2\sigma$ of 
the distribution of $\beta_c$.
We can see from Fig.~\ref{fig:betac_Ising} that $\beta_c(L)$ 
rapidly approaches the exact value $\ln(1+\sqrt{2})/2=0.4406868$, 
which is denoted by a dotted line, even for small sizes.  
The rapid convergence of the present algorithm is apparent 
when we compare the size-convergence rate of $\beta_c(L)$ 
with that of the original PCC (Fig.~1 of Ref.~\cite{PCC}). 
In the original version of the PCC algorithm, 
although the size-dependent $\beta_c(L)$ is automatically tuned, 
we still have to consider the size dependence of $\beta_c(L)$ 
based on the FSS.  Instead, with the two-size PCC algorithm, 
the infinite-size critical temperature is easily achieved 
even for small sizes.

\begin{figure}
\begin{center}
\includegraphics[width=8.6cm]{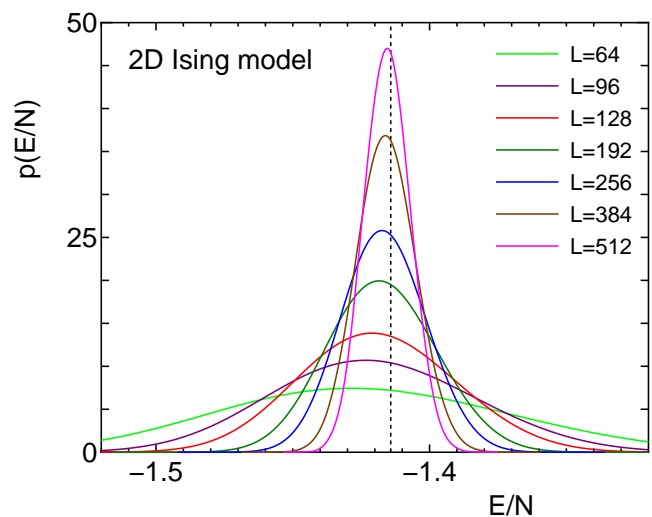}
\caption{
The plot of $p(E/N)$ for the 2D Ising model. 
The system sizes are $L$ = 64, 96, 128, 192, 256, 384, and 512; 
the condition for averaging is
$N_{\rm av}$=4000, $\Delta \beta$=0.00005.
}
\label{fig:p(E)i}
\end{center}
\end{figure}

The energy distribution $p(E/N)$ 
is plotted in Fig.~\ref{fig:p(E)i} 
for $L$ = 64, 96, 128, 192, 256, 384, and 512. 
The energy distribution is normalized by
$$
   \int p(E/N) \ d(E/N) = 1.
$$
There is a single peak, which will be compared with 
the case of the first-order transition later. 
We see that the distribution becomes sharper as the system size increases. 
The peak position approaches the exact energy at the critical temperature, 
that is, $E/N$ = $-\sqrt{2}=-1.41421$.  The exact critical value  
is denoted by a dotted line in the figure. 

\begin{figure}
\begin{center}
\includegraphics[width=8.6cm]{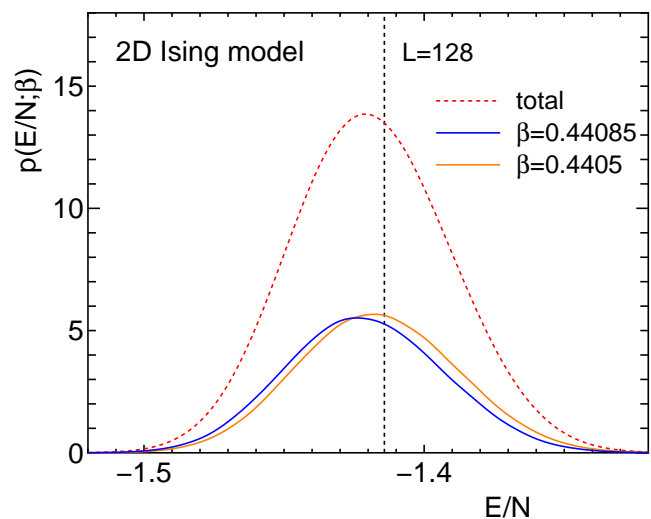}
\caption{
The plot of the $\beta$-decomposed energy distribution $p(E/N;\beta)$ 
for the 2D Ising model. 
The system size is $L=128$. 
The data of two $\beta$'s are compared. 
The whole distribution $p(E/N)$ is shown by a dotted line. 
}
\label{fig:p(E)beta_i}
\end{center}
\end{figure}

Because each system wanders around temperature, 
we can take a thermal average of physical quantities 
at each temperature. This is the same situation 
encountered in the replica exchange method \cite{Hukusima}. 
There, the temperatures of replicas are exchanged 
following the transition probabilities 
based on the Boltzmann weight; 
the thermal average at a fixed $\beta$ is obtained 
by averaging over different replicas. 
Before showing the data of the thermal average of physical quantities, 
we present the energy distribution for a fixed value of $\beta$. 
The energy distribution is decomposed as 
\begin{equation}
  p(E) = \sum_{\beta} p(E;\beta).
\end{equation}
The data of $p(E/N;\beta)$ at two typical temperatures, 
$\beta$ = 0.44085 and 0.4405, 
together with the whole distribution of $p(E/N)$, 
is shown in Fig.~\ref{fig:p(E)beta_i}. 
The system size is fixed at $L=128$. 
The value of the $\beta$-decomposed distribution is 
magnified twenty times for clarity. 
Two energy distributions with different values of $\beta$ are related 
to each other through the equation 
\begin{equation}
  p(E;\beta') \propto e^{-(\beta' - \beta)E} \ p(E;\beta) .
\end{equation}
It is a reweighting of the Boltzmann factor, which is the basis of 
the histogram method by Ferrenberg and Swendsen \cite{Ferrenberg88}. 
The thermal average of a physical quantity $A$ at $\beta'$ 
is obtained by the measurement at $\beta$ through the relation 
\begin{equation}
  \l A \r_{\beta'} = \frac{[A(E) \ e^{-(\beta'-\beta)E}]_{\beta}}
                  {[e^{-(\beta'-\beta)E}]_{\beta}},
\end{equation}
where $[ \cdots ]$ stands for the Monte Carlo average at $\beta$. 
We note that the $\beta$-decomposed energy distribution, 
$p(E/N; \beta)$, does not depend on $h(\beta)$.  
Thus, we obtain the thermal average of a physical quantity $A$, 
$\l A \r_{\beta}$, without considering 
$N_{\rm av}$ and $\Delta \beta$.

\begin{figure}
\begin{center}
\includegraphics[width=8.6cm]{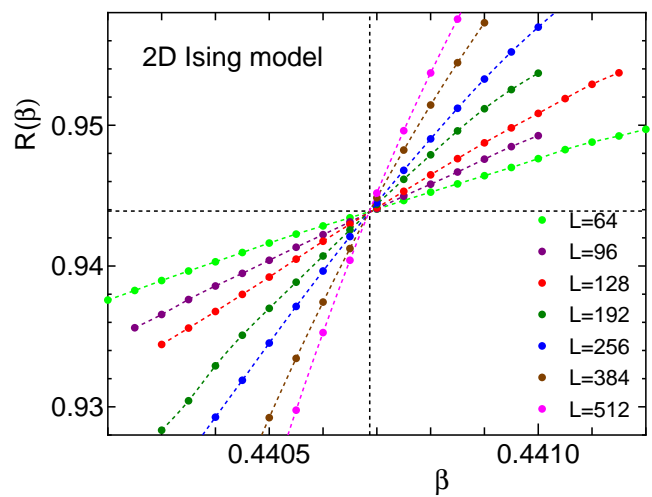}
\caption{
The plot of $R(\beta)$ for the 2D Ising model. 
The system sizes are $L$ = 64, 96, 128, 192, 256, 384, and 512. 
The exact values of $\beta_c$ and $R_c$ are given by a dotted line. 
}
\label{fig:R(E)i}
\end{center}
\end{figure}

As an example of the physical quantities, we show 
the correlation ratio $R(\beta)$ 
as a function of temperature for various system sizes 
in Fig.~\ref{fig:R(E)i}. We see that the data of different 
sizes intersect at the critical point within statistical errors. 
The critical value of the correlation ratio, $R_c$, 
is calculated as follows:
\begin{eqnarray}
 \frac{\l g(L/2) \r}{\l g(L/4) \r} &=& 
 \frac{\displaystyle |\theta_1(1/2)|^{-1/4} \ 
       \sum_{\nu=1}^4 |\theta_{\nu}(1/4)|}
      {\displaystyle |\theta_1(1/4)|^{-1/4} \ 
       \sum_{\nu=1}^4 |\theta_{\nu}(1/8)|} 
      \nonumber \\
 &=& 0.943905, 
\end{eqnarray}
using the Jacobi $\theta$-functions \cite{Francesco} 
(see also Refs.~\cite{Salas2000,tomita2002b}). 
This value is denoted by a dotted line 
in Fig.~\ref{fig:R(E)i}. 
Our simulation reproduces the exact value of $R_c$ 
with an accuracy up to four digits.

\begin{figure}
\begin{center}
\includegraphics[width=8.6cm]{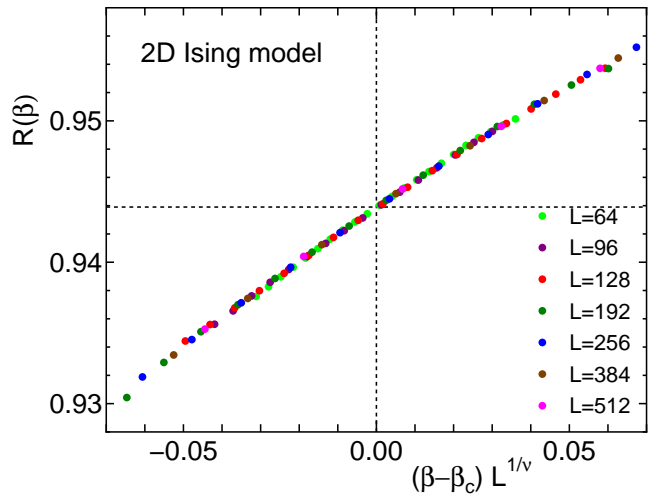}
\caption{
The FSS plot of $R(\beta)$ for the 2D Ising model. 
The system sizes are $L$ = 64, 96, 128, 192, 256, 384, and 512. 
}
\label{fig:R(b)scalei}
\end{center}
\end{figure}

Now let us consider the FSS. 
Because the critical temperature and the critical exponents are 
known for the 2D Ising model, $R(\beta)$ 
are plotted as a function of $(\beta-\beta_c)L^{1/\nu}$, 
as shown in Fig.~\ref{fig:R(b)scalei}, 
where $\beta_c=\ln(1+\sqrt{2})/2$ and $1/\nu=1$. 
We see that the FSS works quite well. 


\section{First-order phase transition}

We now consider the case of the first-order phase transition. 
The 2D ferromagnetic $q$-state Potts model 
\cite{Potts,Kihara,Wu} is taken into account. 
The Hamiltonian is given by
\begin{equation}
 H = J \sum_{\l ij \r} (1 - \delta_{s_i s_j}), \quad s_i = 1, 2, \cdots, q, 
\label{Potts}
\end{equation}
where $\delta_{a b}$ is the Kronecker delta. 
This model is known to show the second-order phase transition 
for $q \le 4$ and first-order phase transition 
for $q \ge 5$. 

Here, we provide the data for a two-size PCC calculation 
of the 2D 6-state Potts model. 
Hysteresis in the first-order transition systems should be considered, 
which is different from the conditions  of 
the second-order transition. It is more feasible to employ 
the multi-cluster update of the Swendsen-Wang type \cite{sw87} because 
a spin configuration changes extensively with such an update. 
For the systems with the second-order transition, there is 
no appreciable difference in the choice of the cluster update. 
The number of the steps for calculating the short-time average, 
$N_{\rm av}$, and the difference in $\beta$, $\Delta \beta$, 
were chosen as 4000 and 0.00001, respectively. We used smaller 
values of $\Delta \beta$ as this could help reduce the effect of hysteresis. 
We conducted measurements for $4 \times 10^6$ steps after equilibration 
of 10,000 steps; such measurements were repeated 64 times 
for $L$ = 64 and 96, and 32 times for $L$ = 128, 192, 256, 384, and 512. 
We note that the correlation function of the $q$-state 
Potts model is given as
\begin{equation}
  g(r) = \frac{q \sum_i \delta_{s_i s_{i+r}} - N}{q-1}.
\end{equation}

\begin{figure}
\begin{center}
\includegraphics[width=8.6cm]{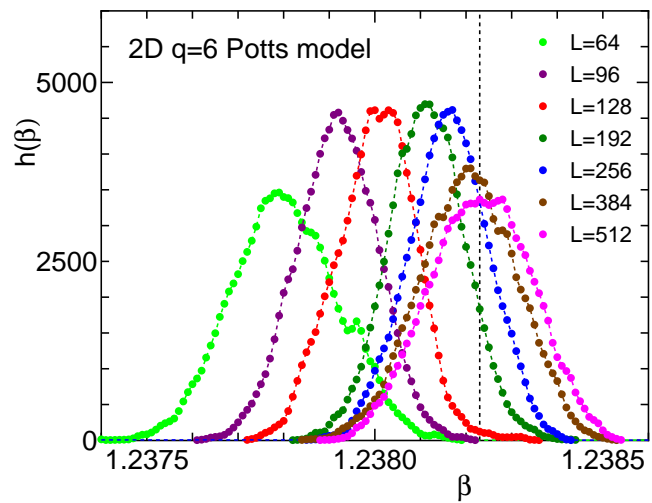}
\caption{
The plot of $h(\beta)$ for the 2D $q=6$ Potts model. 
The system sizes are $L$ = 64, 96, 128, 192, 256, 384, and 512; 
the condition for averaging is
$\Delta \beta$=0.00001, $N_{\rm av}$=4000.
The exact value of $\beta_c$ (=$\ln(1+\sqrt{6})=1.23823$) 
for the infinite system is given by a dotted line.
}
\label{fig:h(b)p}
\end{center}
\end{figure}

The histogram of $\beta$, $h(\beta)$, for the system sizes 
$L$ = 64, 96, 128, 192, 256, 384, and 512 is shown in Fig.~\ref{fig:h(b)p}. 
The histogram exhibits sharp peaks, and the peak position gradually 
approaches the exact value. 
The exact value of the first-order transition inverse 
temperature for the infinite system is given by 
$\ln(1 + \sqrt{q})$ = 1.23823 (for $q=6$), in units of $J$, 
which is shown by the dotted line.

\begin{figure}
\begin{center}
\includegraphics[width=8.6cm]{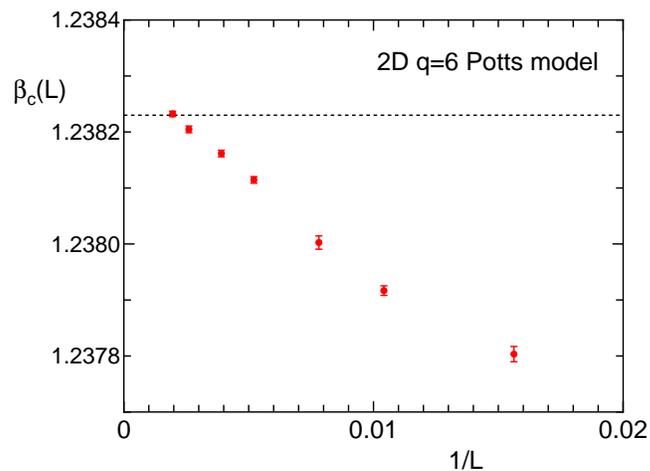}
\caption{
The plot of $\beta_c(L)$ 
of the 2D $q=6$ Potts model. 
The linear system sizes $L$ are 64, 96, 128, 192, 256, 384, and 512. 
The exact value is denoted by a dotted line.
}
\label{fig:betac_Potts}
\end{center}
\end{figure}

The average value of $\beta$, $\beta_c(L)$, is plotted 
as a function of $1/L$ in Fig.~\ref{fig:betac_Potts}. The exact value of 
the first-order transition inverse temperature (1.23823) 
is given by the dotted line. We can observe that the 
calculated estimate of the transition temperature 
converges to the exact value with five-digit accuracy. 

\begin{figure}
\begin{center}
\includegraphics[width=8.6cm]{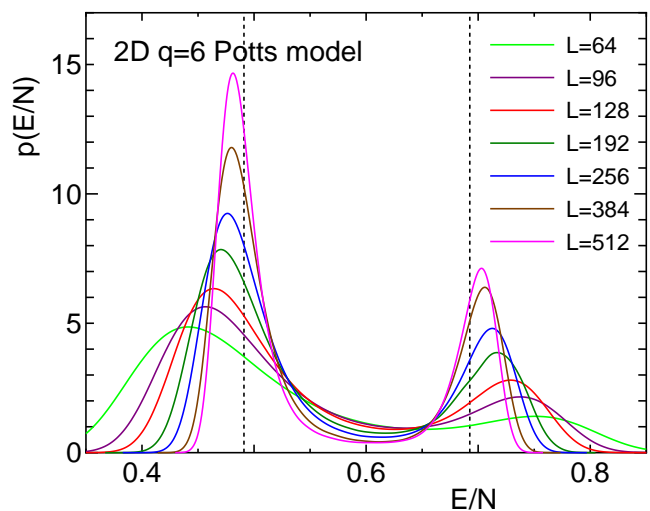}
\caption{
The plot of $p(E/N)$ for the 2D $q=6$ Potts model. 
The system sizes are $L$ = 64, 96, 128, 192, 256, 384, and 512. 
Baxter's results \cite{Baxter} for two peak positions 
are given by dotted lines. 
}
\label{fig:p(E)p}
\end{center}
\end{figure}

\begin{figure}
\begin{center}
\includegraphics[width=8.6cm]{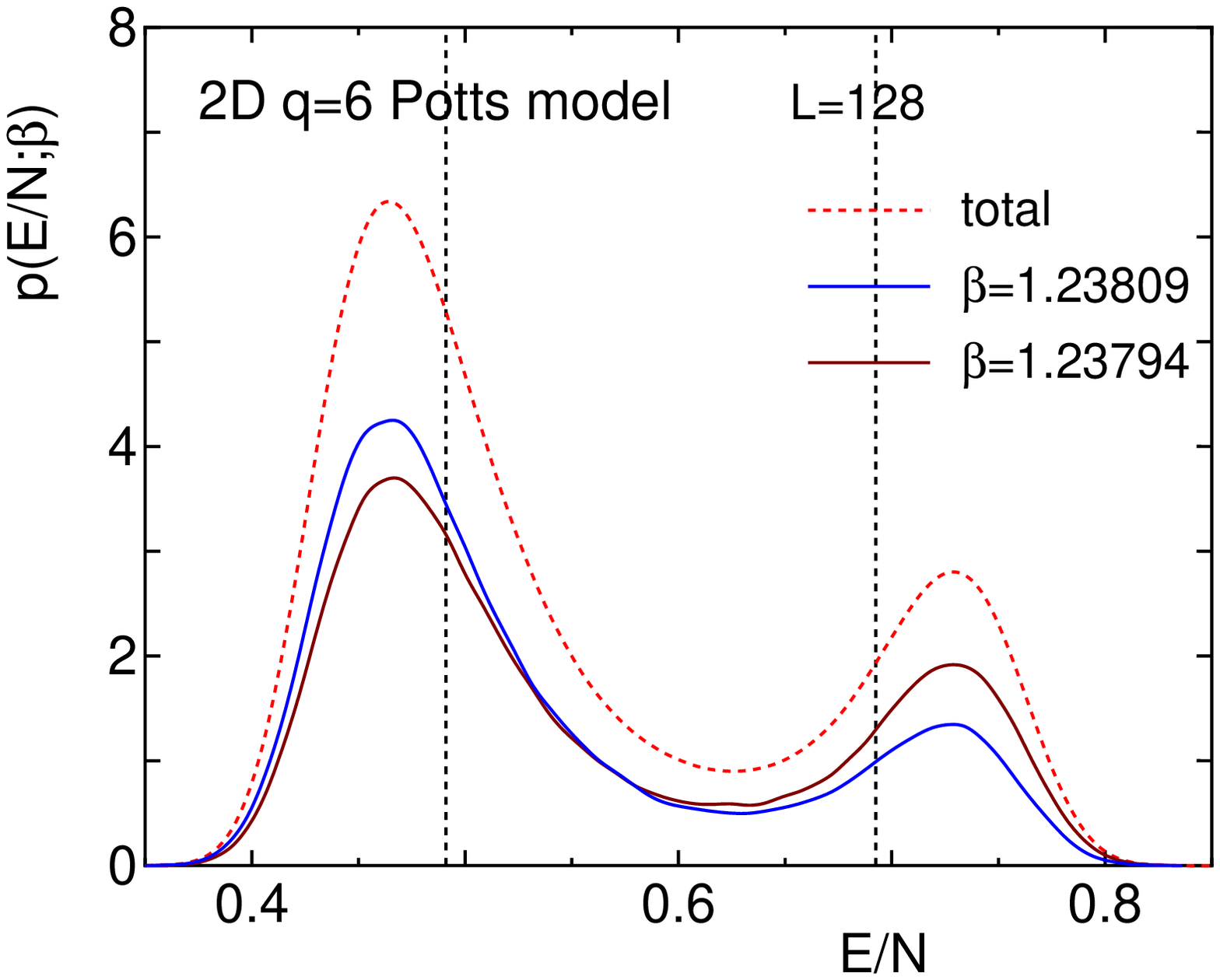}
\caption{
The plot of the $\beta$-decomposed energy distribution $p(E/N;\beta)$ 
for the 2D $q=6$ Potts model. 
The system size is $L=128$. 
The data of two $\beta$'s are compared. 
The whole distribution $p(E/N)$ is shown by a dotted line. 
}
\label{fig:p(E)beta_p}
\end{center}
\end{figure}

The distribution of $E$, $p(E/N)$, is shown in Fig.~\ref{fig:p(E)p} 
for various sizes. 
We observe double peaks, which are specific to the first-order transition. 
In Fig.~\ref{fig:p(E)beta_p}, we plot the $\beta$-decomposed 
energy distribution, $p(E/N; \beta)$.  The system size is set to be $L=128$. 
Here, we show the data of two typical temperatures, 
$\beta$ = 1.23809 and 1.23794, which are on both sides of 
the peak value of $h(\beta)$ shown in Fig.~\ref{fig:h(b)p}, 
together with the entire distribution of $p(E/N)$. 
The value of the $\beta$-decomposed distribution is 
magnified twenty times for clarity. 
We can observe that the weight of high energies increases 
for the high-temperature (low-$\beta$) energy distribution. 

\begin{figure}
\begin{center}
\includegraphics[width=8.6cm]{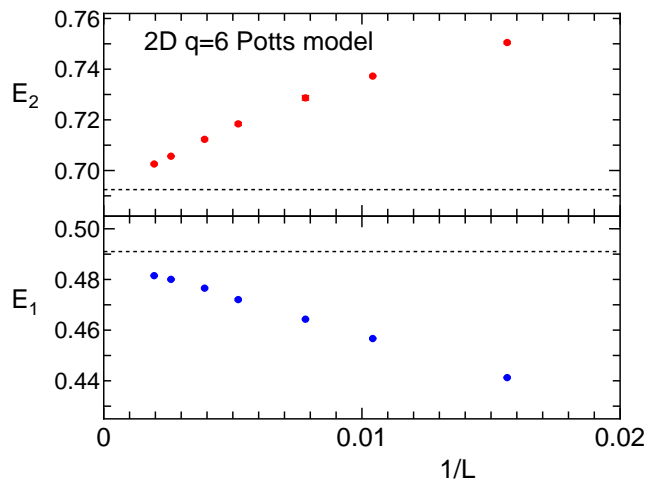}
\caption{
The plot of $E_1$ and $E_2$ as a function of $1/L$ 
for the 2D $q=6$ Potts model. 
The system sizes are $L$=64, 96, 128, 192, 256, 384, and 512. 
Baxter's results \cite{Baxter} for two peak positions 
are given by dotted lines. 
}
\label{fig:E12}
\end{center}
\end{figure}

We now examine the peak positions of energy. 
Baxter \cite{Baxter} (see also \cite{Binder81b}) 
calculated the exact difference in the higher energy peak $E_2$ and 
the lower energy peak $E_1$, i.e., the latent heat. 
The exact result is 
\begin{equation}
   (E_2- E_1)/N = 2 ( 1+\frac{1}{\sqrt{q}}) \tanh \frac{\Theta}{2} 
   \prod_{n=1}^{\infty} (\tanh n\Theta )^2,
\end{equation}
where $\Theta = \textrm{arcosh} (\sqrt{q}/2)$. 
The middle point $(E_1+E_2)/2N$ is also given as
\begin{equation}
   (E_1+E_2)/2N = 1-1/\sqrt{q}. 
\end{equation}
Thus, for $q=6$, $E_1/N$ and $E_2/N$ are calculated 
as 0.49102 and 0.69248, respectively. 
These values are given in Fig.~\ref{fig:p(E)p}.  
We can observe that the positions of the energy peaks 
approach the exact infinite values as the system size increases.
The size dependences of the numerical estimates of $E_1/N$ and $E_2/N$ 
are plotted as a function of $1/L$ in Fig.~\ref{fig:E12}. 
The statistical errors are within the size of marks. 
They converge to the exact values \cite{Baxter}.

\section{BKT transition}
The 2D spin systems with continuous XY symmetry exhibit a unique 
phase transition called the Berezinskii-Kosterlitz-Thouless (BKT) 
transition \cite{Berezinskii,kosterlitz,kosterlitz2}. 
There exists a BKT phase of a quasi long-range order (QLRO), 
where the correlation function decays as a power law. 
Here, we consider the $q$-state clock model, which is a discrete version 
of the classical XY model. 
The Hamiltonian is given by
\begin{equation}
   H = -J \sum_{\l ij \r} \cos (\theta_i - \theta_j), \quad 
   \theta_i = 2\pi i/q, \ i = 1, 2, \cdots, q.
\end{equation}
The 2D $q$-state clock model experiences the BKT transition 
for $q \ge 5$, whereas the $q=4$ clock model is 
two sets of the Ising model and the 3-state clock model 
is equivalent to the 3-state Potts model. 
The $q=2$ clock model is simply the Ising model.

For $q \ge 5$, there is an interplay between the plane-rotator symmetry,
which attempts to preserve the BKT phase, and the discreteness, 
which tends to create a long-range order (LRO) at low temperatures.
Two transition temperatures, $T_1 < T_2$, are observed; 
each corresponds to the transition between LRO and QLRO and 
between QLRO and a disordered phase. 

\begin{figure}
\begin{center}
\includegraphics[width=8.6cm]{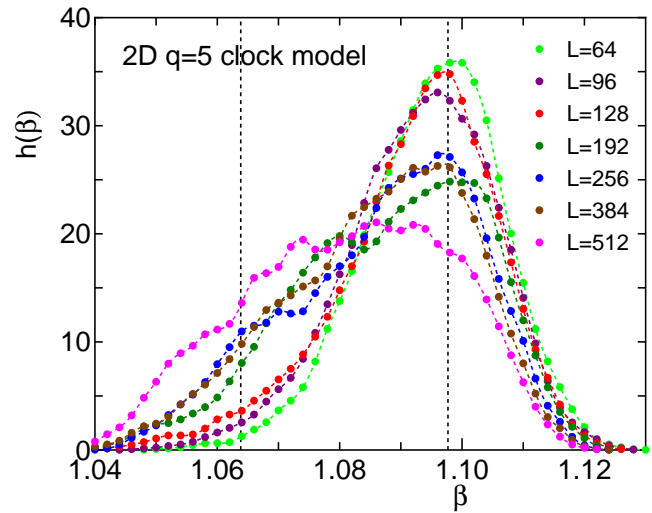}
\caption{
The plot of $h(\beta)$ for the 2D $q=5$ clock model. 
The system sizes are $L$ = 64, 96, 128, 192, 256, 384, and 512. 
The numerical estimates of $\beta_1$ and $\beta_2$ are given by dotted lines.
}
\label{fig:h(b)c}
\end{center}
\end{figure}

We conducted a simulation of the two-size PCC algorithm for the 2D $q=5$ 
clock model.  As the system has a wide temperature range in 
the critical state, we selected a larger $\Delta \beta$, 0.002. 
Again, we selected $N_{\rm av}$=4000. 
A histogram of $\beta$, $h(\beta)$, of the $q=5$ clock model is shown 
in Fig.~\ref{fig:h(b)c}.  The data in the histogram are widely distributed, 
which is contrast to the case of strong transitions, i.e., 
the second-order and first-order transitions. 
This is related to the fact that there is a fixed line 
instead of a fixed point for the system with the BKT transition. 
In the figure, the numerical estimates of $\beta_1$ ($1/T_1 = 1/0.911(5)$) 
and $\beta_2$ ($1/T_2 = 1/0.940(5)$) \cite{Surungan} are shown 
by dotted lines for convenience. 

\begin{figure}
\begin{center}
\includegraphics[width=8.6cm]{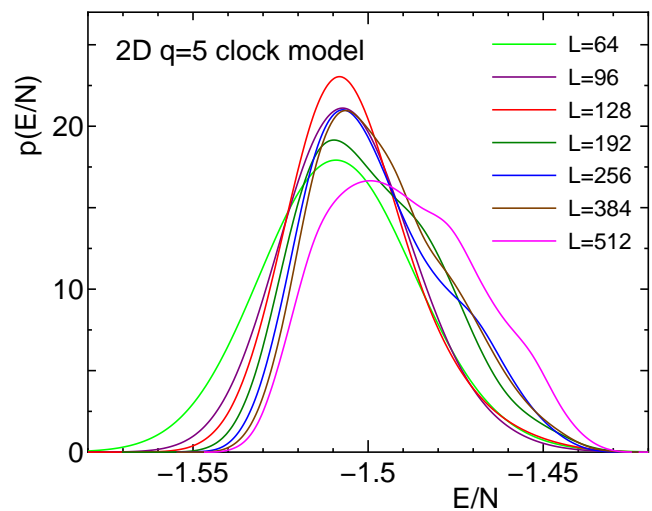}
\caption{
The plot of $p(E)$ for the 2D $q=5$ clock model. 
The system sizes are $L$ = 64, 96, 128, 192, 256, 384, and 512. 
}
\label{fig:p(E)c}
\end{center}
\end{figure}

\begin{figure}
\begin{center}
\includegraphics[width=8.6cm]{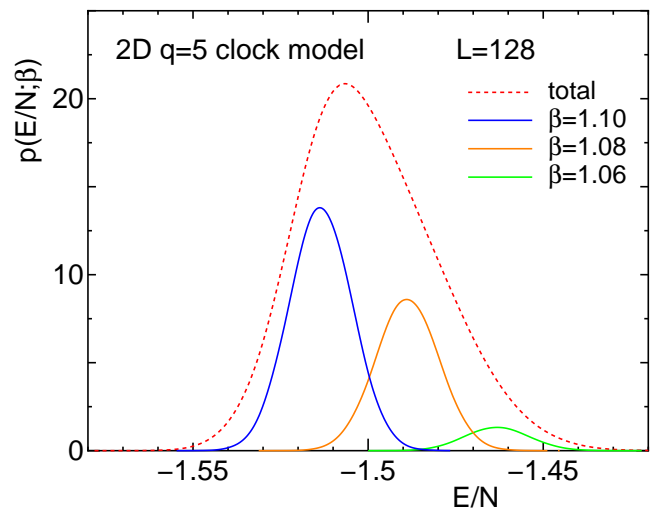}
\caption{
The plot of the $\beta$-decomposed energy distribution $p(E/N;\beta)$ 
for the 2D $q=5$ clock model. 
The system size is $L=128$. 
We show the data for three $\beta$'s; from left, $\beta$=1.10, 1.08, 
and 1.06.
The whole distribution $p(E/N)$ is shown by a dotted line. 
}
\label{fig:p(E)beta_c}
\end{center}
\end{figure}

The distribution of $E$, $p(E/N)$, is shown in Fig.~\ref{fig:p(E)c} 
for various sizes. 
In Fig.~\ref{fig:p(E)beta_c}, we plot the $\beta$-decomposed 
energy distribution $p(E/N;\beta)$. The system size is $L=128$. 
Data for three $\beta$'s are shown; from left, $\beta$=1.10, 
1.08, and 1.06.  The value of the $\beta$-decomposed distribution is 
magnified ten times for clarity. 
Energy peaks are observed at certain values depending on the temperature. 

\begin{figure}
\begin{center}
\includegraphics[width=8.6cm]{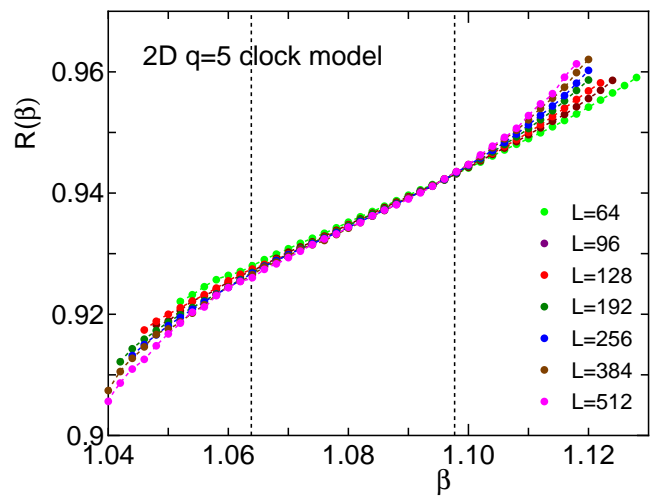}
\caption{
The plot of $R(\beta)$ for the 2D $q=5$ clock model. 
The system sizes are $L$ = 64, 96, 128, 192, 256, 384, and 512. 
The numerical estimates of $\beta_1$ and $\beta_2$ are given by dotted lines.
}
\label{fig:R(b)c}
\end{center}
\end{figure}

We calculate the temperature dependence of physical quantities 
using the same procedure as the one followed by the Ising model. 
Although the histogram $h(\beta)$ shown in Fig.~\ref{fig:h(b)c} 
is not very smooth, fairly accurate estimates of the thermal average 
of physical quantities at a fixed $\beta$ can be obtained 
as was discussed in the case of the Ising model. 
The temperature dependence of the correlation ratio for various sizes 
is plotted in Fig.~\ref{fig:R(b)c}. The numerical estimates 
of $\beta_1$ and $\beta_2$ reported in \cite{Surungan} are shown  
by dotted lines for convenience. 
In the intermediate temperature range, the correlation ratios 
of different sizes take the same value, whereas they start 
to exhibit variations below $\beta_2$ and above $\beta_1$.  
To locate the BKT transition 
temperatures precisely, a careful FSS treatment with exponential 
divergence behavior is required \cite{Surungan}. 
When the present method is used directly, 
the systems remain in the intermediate state for a long time. 
We may set windows for the allowed temperature range. 
In the case of the 2D $q=5$ clock model, the temperature range 
may, for example, be restricted as $\beta < 1.07$ for the $\beta_2$ transition, 
and $\beta>1.09$ for the $\beta_1$ transition. 

\section{Summary and discussion}

In this paper, we described the two-size PCC algorithm. 
We simultaneously simulate two systems of different sizes 
at the same temperature. Comparing the short-time average of 
the correlation ratios of the two sizes, we increase or decrease 
the temperature based on the negative feedback mechanism 
given by Eq.~(\ref{feedback}).  
A temperature near the critical temperature 
is automatically selected. 
It is simply an Ehrenfest model for 
{\it diffusion with a central force} \cite{Ehrenfest,Feller}.

For the strong transitions including second-order 
or first-order transitions, the temperature peaks sharply 
at the critical temperature. Thus, we can locate the critical temperature 
in a self-adapted manner.  The energy distribution is 
singly peaked in the case of the second-order transition, 
whereas it is doubly peaked in the case of the first-order transition.  
As the system wanders around the temperature, we can calculate 
the thermal average of physical quantities for each temperature. 
We showed the results of the correlation ratios 
of the 2D Ising model, which demonstrated a satisfactory FSS behavior. 
For the first-order transition, we determined 
the double-peak positions $E_1$ and $E_2$ for the 2D $q=6$ 
Potts model. The results were compared with the exact values 
obtained by Baxter~\cite{Baxter}. 

In the case of the systems with the BKT transition, 
the temperature is widely distributed in the two-size 
PCC algorithm, which is owing to the existence 
of a fixed line in the BKT transition. 
An investigation of the temperature dependence of the correlation ratio 
for the 2D $q=5$ clock model showed that 
correlation ratios of different sizes 
take the same value in the intermediate BKT state, 
whereas they start to vary below $\beta_2$ and above $\beta_1$.  
We can easily determine the specific behavior of the BKT transition 
compared to the second-order transition or the first-order transition. 

The advantage of the present method is that a temperature 
range can be automatically selected near the critical temperature. 
As the sampling of the temperature peaks at the 
critical temperature, the critical phenomena can be studied 
efficiently. 

To summarize, we have proposed a unified method of numerical simulation 
that can treat the second-order phase transition, the first-order 
phase transition, and the BKT transition with equal footing. 
By simultaneously simulating two systems of different sizes, 
say $L$ and $L/2$, we could measure the correlation functions, 
which are essential when investigating phase transition. 
Thus, we could easily determine the type of the phase transition. 
The proposed algorithm is general.  We can apply this algorithm 
to various problems of any dimension. 
For example, the 2D ferromagnetic $q$-state Potts model 
with $r$ invisible (redundant) states exhibits a change in 
the phase transition from the second order to the first order 
owing to the entropy effect of invisible states \cite{tamura10}. 
A study on the two-size PCC algorithm is now in progress for such 
a transition change problem. 

\vspace*{2mm}

\section*{Acknowledgments}

The authors wish to thank Yukihiro Komura 
for the collaboration in the early stage of research. 
The HPC facilities of the Indonesian Institute of Science 
(LIPI) were used for computation. 
This work was supported by a Grant-in-Aid for Scientific Research 
from the Japan Society for the Promotion of Science, Grant Number JP16K05480.
TS is grateful to Fundamental Research Grant 
from Hasanuddin University, FY 2019.

\end{document}